\documentclass[aps,prb,amsfonts,amssymb,amsmath,groupedaddress,twocolumn]{revtex4-2}

\usepackage{graphicx}
\usepackage{amsmath}
\usepackage{amsfonts}
\usepackage{amssymb}
\usepackage{graphicx}
\usepackage{epstopdf}

\begin{document}

\title{First-principles analysis of intravalley and intervalley electron-phonon scattering in thermoelectric materials}

\author{Vahid Askarpour}
\affiliation{Department of Physics and Atmospheric Science, Dalhousie University, Halifax, Nova Scotia, Canada, B3H 4R2}
\author{Jesse Maassen}
\email{jmaassen@dal.ca}
\affiliation{Department of Physics and Atmospheric Science, Dalhousie University, Halifax, Nova Scotia, Canada, B3H 4R2}

\begin{abstract}
Intervalley collisions, which scatter electrons from one valley or band to another, can be detrimental to thermoelectric performance in materials with multiple valleys/bands. In this study, density functional theory is used to investigate the electron-phonon scattering characteristics of three lead chalcogenides (PbS, PbSe, PbTe) and three half-Heuslers (ScNiBi, ScPdSb, ZrNiSn), which all possess multiple equivalent conduction valleys, in order to characterize and analyze their intravalley/intervalley components. To elucidate what controls the degree of intravalley and intervalley transitions, the scattering rates are decomposed into the product of the phase space (a measure of how much scattering is possible) and the average electron-phonon coupling. To help guide the search for improved thermoelectric and high-conductivity materials, simple and approximate approaches are demonstrated that can be adopted to identify materials with reduced intervalley scattering, which circumvent the need for computationally-demanding electron-phonon scattering calculations. In addition, the benefits of selecting materials with large-energy zone-edge phonons are explored in the limit $\hbar\omega$\,$\gg$\,$k_BT$, and found to potentially suppress intervalley processes by up to an order of magnitude, leading to a 70\% and 100\% increase in conductivity and power factor, respectively.
\end{abstract}

\maketitle

\section{Introduction} 
\label{sec:introduction}
Materials that efficiently convert heat energy into electrical energy, or vice versa, display a high thermoelectric (TE) figure of merit $ZT$\,=\,$S^2 \sigma T / \kappa$, where $S$ is the Seebeck coefficient, $\sigma$ is the electrical conductivity, $T$ is the temperature and $\kappa$ is the sum of electronic and lattice thermal conductivities \cite{Snyder2008,He2017}. One way to enhance $ZT$ is to optmize the power factor $PF$\,=\,$S^2 \sigma$, which can be challenging since $S$ and $\sigma$ are interrelated -- increasing $|S|$ tends to decrease $\sigma$, and vice versa. Band engineering is a broad strategy to control and modify the electronic states and scattering properties of TE materials with the goal of enhancing the power factor \cite{Pei2012a,Zhao2014,Xin2018}. Such approaches include, for example, band convergence \cite{Lee2020}, resonant states \cite{Heremans2012}, energy filtering \cite{Faleev2008,Bahk2013,Thesberg2016}, modulation doping \cite{Zebarjadi2011,Yu2012,Neophytou2016}, warped electronic bands \cite{Chen2013,Maassen2013,Mori2013,Wickramaratne2015}, and optimization of scattering profiles \cite{Shuai2017,Mao2017}.

Band convergence, in particular, has been the focus of much research over the past decade, and is based on a strategy aligning multiple electronic valleys and/or bands within a narrow energy range (compared to $k_BT$). Multiple valleys can arise from valley degeneracy and/or secondary higher-energy bands that may themselves be degenerate. Control over band alignment can come from alloying, strain and temperature, each of which can shift the relative energy of the band edges. Aligning multiple valleys/bands can enable a higher $|S|$ with a similar $\sigma$, when assuming a fixed carrier concentration, thereby enhancing the power factor. The effect on the TE parameters depends on a number factors, including the shape of the bands, number of valleys and scattering rates.

Improved TE performance by band convergence has been illustrated in many cases. Pei et al. experimentally demonstrated $ZT$ of 1.8 in p-PbTe$_{1-x}$Se$_x$ originating from the alignment of multiple valleys tuned through alloying of Se and temperature \cite{Pei2011}. Liu et al. found that n-Mg$_2$Si$_{1-x}$Sn$_x$ exhibits a large power factor and $ZT$ reaching 1.3 when $x$\,=\,0.7 resulting from the convergence of light and heavy conduction bands, confirmed by density functional theory (DFT) and experiment \cite{Liu2012}. Tang et al. explained, through a combination of measurements and DFT, how the high $ZT$ of the skutterudite n-CoSb$_3$ stems from the convergence of a single primary conduction band with highly degenerate secondary bands \cite{Tang2015}. Zhang et al. combined DFT and experiment to show that the high TE performance in the Zintl phase n-Mg$_3$Sb$_{1.5}$Bi$_{0.5}$ arises from multi-valley conduction leading to an enhanced power factor \cite{Zhang2017a}. Many other studies confirming the benefits of band convergence have been reported \cite{Zhu2014,Wang2016,Kim2017,Zheng2017,Zheng2017,Li2017,Li2018,Sun2018,Diznab2019,Hung2019}.

Although having multiple valleys participate in carrier transport has been shown to benefit TE performance, the degree of improvement will depend sensitively on the scattering details -- in particular, the magnitude of intervalley (or interband) scattering. In materials with a single valley, only intravalley scattering processes occur, wherein the electron (or hole) transitions between initial and final states both residing within the same valley. However, in the case of multiple valleys or bands, carriers can additionally scatter from one valley to another giving rise to intervalley scattering. Intervalley processes activate when aligning multiple valleys and add to the intravalley processes, thereby increasing the overall scattering rates and potentially reducing the conductivity. As a result, intervalley scattering will tend to offset, and in some cases potentially eliminate, the benefits of band convergence. Thus, reducing intervalley collisions can help achieve the maximum benefits of band convergence.

Theoretical studies capturing the effect of intervalley transitions have revealed that the advantage of band convergence can be significantly offset with strong intervalley scattering. DFT calculations of electron-phonon (el-ph) scattering in silicon, among its six equivalent conduction valleys, showed that the intervalley scattering rates are much larger than those from intravalley processes, which negatively impacts its TE performance \cite{Witkoske2017}. A theoretical investigation of Co-based half-Heuslers by Kumarasinghe and Neophytou \cite{Kumarasinghe2019} found that in certain cases the power factor could decrease with band convergence. Using first-principles, Park et al. \cite{Park2021} carried out scattering and transport calculations on CaZn$_{2-x}$Mg$_x$Sb$_2$ Zintl alloys and the full Heusler Sr$_2$SbAu, and determined that intervalley scattering can render band convergence ineffective. DFT-calculated el-ph scattering and transport simulations of various two-dimensional materials showed a large reduction in conduction and TE performance due to intervalley processes \cite{Wu2021a,Wu2021b}. First-principles calculations of TE transport in p-PbTe revealed that band convergence is beneficial for $ZT$, but that intervalley scattering significantly reduces the power factor \cite{DSouza2022}. When theoretically exploring band convergence it is important that the scattering model captures the effect of intervalley processes, by using a rigorous treatment \cite{Witkoske2017,Severin2018,Park2021,Fedorova2022} or having the scattering rates be proportional to the density of states (DOS) \cite{Graziosi2019,Cameron2021} (assumes an equal probability of scattering to any valley) -- a constant relaxation time is blind to changes in scattering as valleys are aligned \cite{Witkoske2017,Wu2021a,Wu2021b}.

These previous studies highlight the importance of understanding what controls intervalley scattering, with the goal of identifying or designing TEs with limited intervalley processes to maximize the benefits of band converge. In this work, we utilize DFT to carry out el-ph scattering and transport calculations on three lead chalcogenides (PbS, PbSe, PbTe) and three half-Heuslers (ScNiBi, ScPdSb, ZrNiSn) to separate and analyze the contributions of intravalley and intervalley scattering, and their impact on TE performance. These materials were selected since they possess multiple equivalent conduction valleys, with no secondary bands within an energy range relevant for transport. The paper is outlined as follows. Section \ref{sec:theory} discusses the theoretical approach and numerical details. The results and their analysis are presented in Section \ref{sec:results}. Section \ref{sec:discussion} explores how to achieve reduced intervalley scattering and proposes a computationally-efficient scheme for estimating intervalley el-ph coupling. Finally, our findings are summarized in Section \ref{sec:conclusion}.

\section{Theoretical Approach}
\label{sec:theory}
\subsection{Scattering and transport theory}
The el-ph momentum scattering rate for each electronic $\mathbf{k}$-state (includes band and spin index) is calculated using \cite{Lundstrom2000,Liao2015,Askarpour2019}:
\begin{align}
\frac{1}{\tau_{\bf k}} = & \frac{2\pi}{\hbar} \sum_{\bf k'} |g({\bf k},{\bf k'})|^2 \big[ n_{\bf q} + 1/2 \pm (f_{\bf k'} - 1/2) \big]  \nonumber \\
 & \delta(E_{\bf k}-E_{\bf k'} \pm \hbar \omega_{\bf q})    \Big( 1-\frac{{\bf v}_{\bf k'}\cdot{\bf v}_{\bf k}} {|{\bf v}_{\bf k'}|{|\bf v}_{\bf k}|}  \Big), \label{eq:itau}
\end{align}
where $g({\bf k},{\bf k'})$ is the el-ph coupling between the initial state $\bf k$ and the final state $\bf k'$\,=\,${\bf k}\pm{\bf q}$, $E_{\bf k}$ is the electron energy, ${\bf v}_{\bf k}$ is the electron velocity, $f_{\bf k}$ is the Fermi-Dirac distribution, $\bf q$ is the phonon wavevector (includes branch index), $n_{\bf q}$ is the Bose-Einstein distribution, and the upper ``$+$'' and lower ``$-$'' correspond to phonon absorption and emission processes, respectively. The el-ph coupling matrix elements include both standard and polar contributions \cite{Giustino2007,Verdi2015,Sjakste2015}. By tracking the magnitude of the phonon wavevector, the scattering rates are separated into intravalley and intervalley components. Mobile carrier screening is applied to the polar components of the el-ph interaction and dynamical matrices using the Thomas-Fermi theory, as described in Refs.~\cite{Song2017,Askarpour2019}.

A solution of the linear Boltzmann transport equation within the relaxation-time approximation provides the following expressions for the electrical conductivity and Seebeck coefficient \cite{Scheidemante2003,Liao2015}:

\begin{align} 
\sigma_{\alpha} = & \frac{e^2}{\Omega} \sum_{\bf k} {\bf v}_{\bf k}^{\alpha}{\bf v}_{\bf k}^{\alpha} \tau_{\bf k} \Big[ -\frac{\partial f_{\bf k}}{\partial E_{\bf k}} \Big],   \label{eq:sigma} \\
 S_{\alpha} = & \frac{-e}{\sigma_{\alpha} T {\Omega}}  \sum_{\bf k} {\bf v}_{\bf k}^{\alpha} {\bf v}_{\bf k}^{\alpha} {\tau}_{\bf k} (E_{\bf k}-\mu) \Big[ -\frac{\partial f_{\bf k}}{\partial E_{\bf k}} \Big],   \label{eq:seebeck}
\end{align}
where $\mu$ is the chemical potential, $e$ is the electron charge magnitude and $\Omega$ is the sample volume. The power factor is obtained from $PF_{\alpha}$\,=\,$S^2_{\alpha}\sigma_{\alpha}$.

\begin{figure*}[ht]
\includegraphics[width=6.5in]{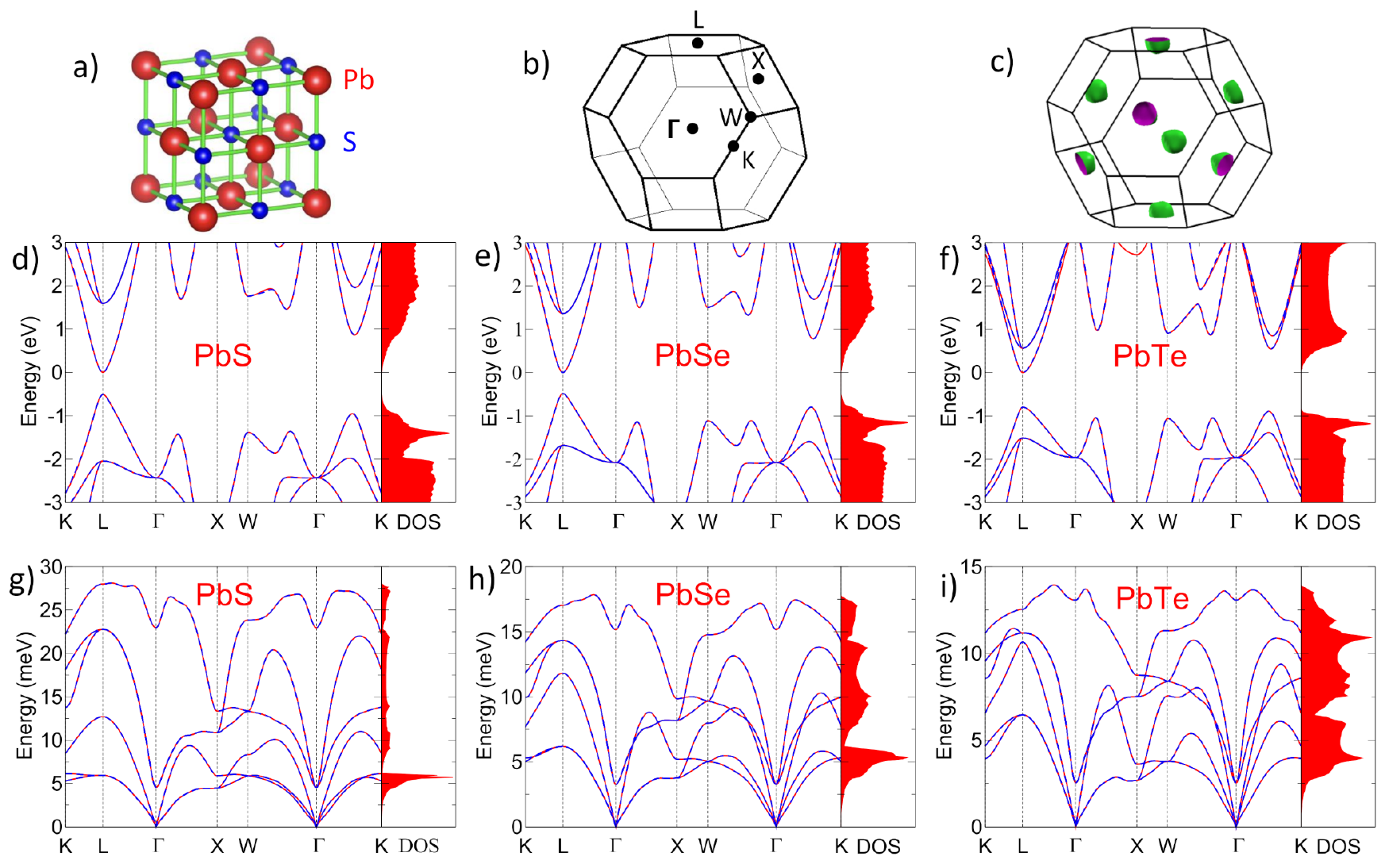}
\caption{a) Unit cell and atomic structure of PbS (shared with PbSe and PbTe). b) Brillouin zone of PbX materials. c) Fermi surface of PbS for an energy of 0.4\,eV above the conduction band edge. d)-f) Electron dispersion and density of states. g)-i) Phonon dispersion and density of states. Red lines are obtained from QE and blue dashed lines are obtained from EPW.} \label{fig:bands_PbX}
\end{figure*}

\begin{figure*}[ht]
\includegraphics[width=6.5in]{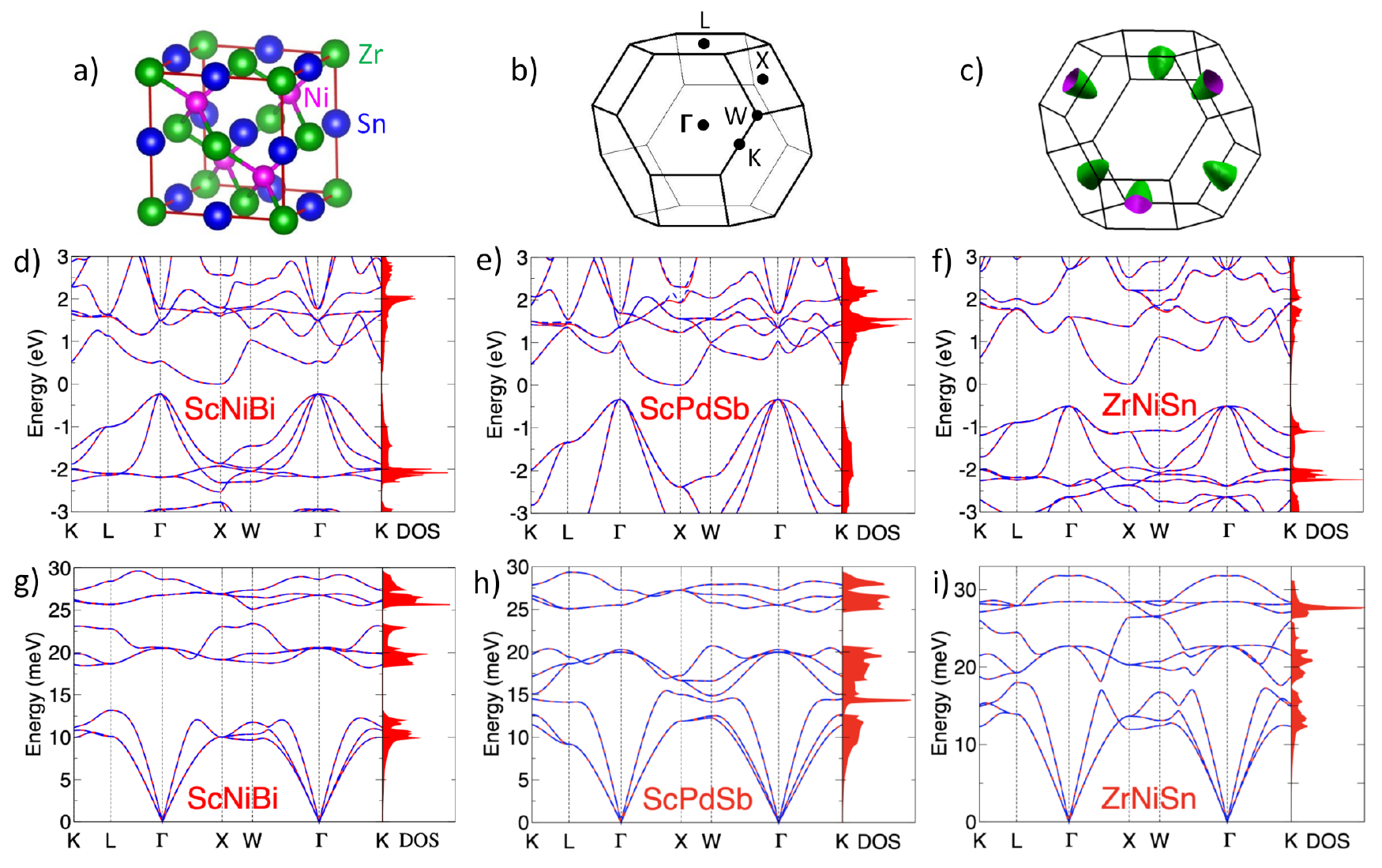}
\caption{a) Unit cell and atomic structure of ZrNiSn (shared with ScNiBi and ScPdSb). b) Brillouin zone of HH materials. c) Fermi surface of ZrNiSn for an energy of 0.2\,eV above the conduction band edge. d)-f) Electron dispersion and density of states. g)-i) Phonon dispersion and density of states. Red lines are obtained from QE and blue dashed lines are obtained from EPW.} \label{fig:bands_HH}
\end{figure*}

\subsection{Numerical details}
The DFT simulations were carried out using the Quantum Espresso (QE) code \cite{QE1,QE2}. Norm-conserving pseudopotentials with Perdew-Burke-Ernzerhof (PBE) exchange-correlation potential \cite{PBE} were adopted, along with a wavefunction energy cutoff of 120\,Ry and a 8$\times$8$\times$8 $\bf k$-grid. Spin-orbit coupling has a small effect on the electronic band structure the half-Heuslers \cite{Zhou2018}, and was not included in the analysis to avoid splitting of the conduction states. The dynamical matrices, phonon energies and the variation in self-consistent potential due to phonons were obtained with density functional perturbation theory (DFPT) as implemented in the QE code.

The el-ph scattering and transport calculations were performed with the EPW code \cite{Noffsinger2010,Ponce2016}, which was modified to separate the intravalley and intervalley processes. The electronic and phonon properties, along with the el-ph coupling, were evaluated on coarse $\bf k$- and $\bf q$-grids of 8$\times$8$\times$8. The lead chalcogenides (referred to as PbX) and the half-Heuslers (HH) used 13 and 32 maximally localized Wannier functions with s, p, and d symmetries, respectively, generated with the Wannier code \cite{Mostofi2014}. For the scattering rates and TE parameters, fine $\bf k$- and $\bf q$-grids of 200$\times$200$\times$200 were adopted for the PbX materials and 120$\times$120$\times$120 were used for the HH materials, along with a Gaussian broadening of 5\,meV.

\section{Results} 
\label{sec:results}
\subsection{Atomic structure} 
The lead chalcogenides, PbX (X\,=\,S, Se, Te), crystallize in a rock-salt structure (Fm$\bar{3}$m) that contains a basis of two atoms in the primitive cell, as shown in Fig.~\ref{fig:bands_PbX}a). The optimized lattice constants for PbS, PbSe and PbTe are 6.006\,\AA, 6.223\,\AA, and 6.551\,\AA, respectively, which are within 2\% of the experimental values of 5.936\,\AA, 6.121\,\AA, and 6.454\,\AA~\cite{Pei2012b}. The HH crystal structure (F$\bar{4}$3m) consists of a face-centered cubic (FCC) lattice with a basis of three atoms, in which two atoms are transition or rare-earth metals, as depicted in Fig.~\ref{fig:bands_HH}a). The calculated lattice constants for ScNiBi, ScPdSb and ZrNiSn are found to be 6.289\,\AA, 6.421\,\AA, and 6.184\,\AA, respectively, which are within 2\% of the corresponding experimental values of 6.179\,\AA, 6.312\,\AA, and 6.095\,\AA~\cite{Deng2017,Oestreich2003,Gong2019}.

\begin{figure*}[ht]
\includegraphics[width=6.5in]{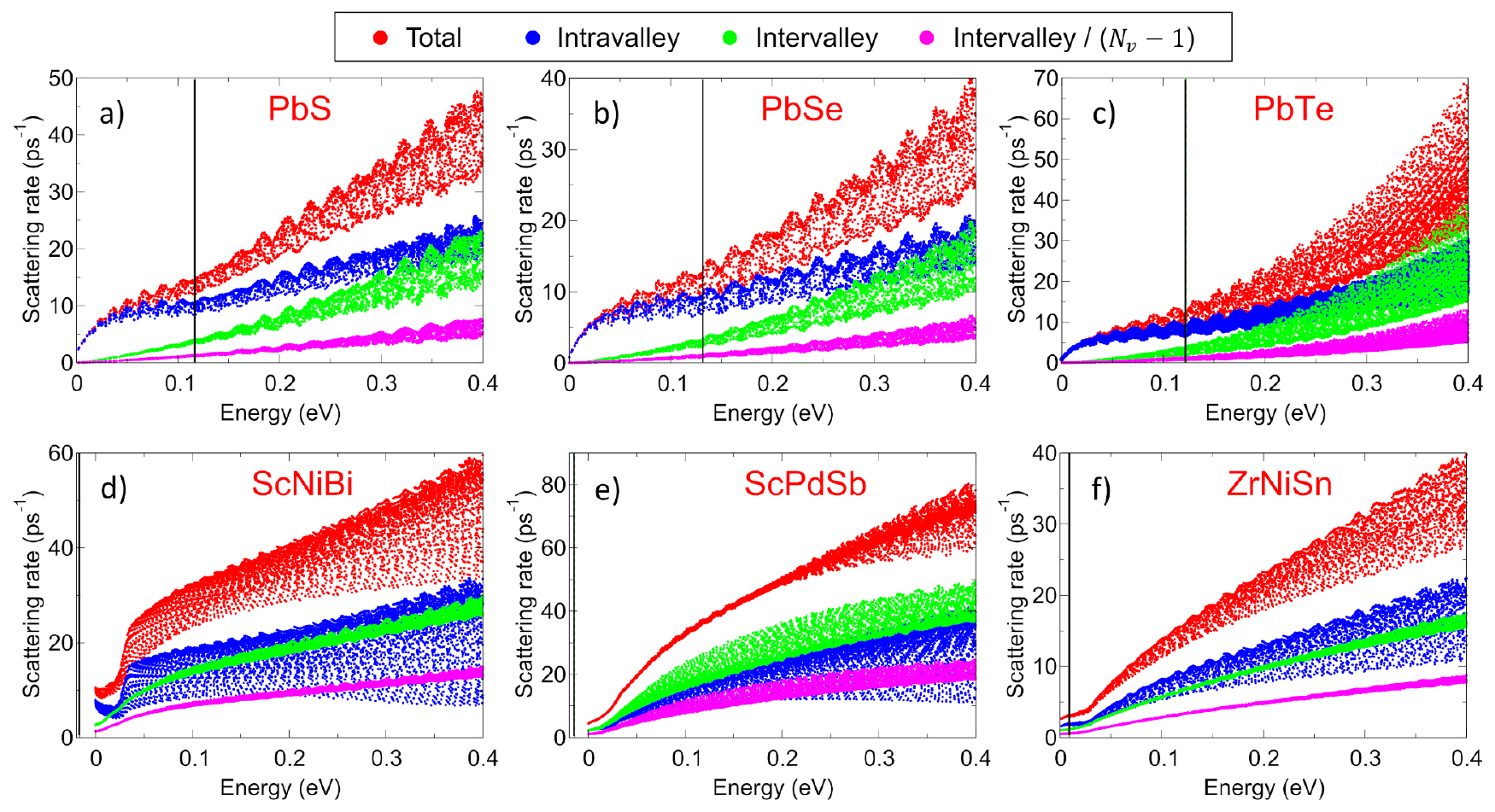}
\caption{Electron-phonon momentum scattering rates, $1/\tau_{\bf k}$, separated into total (red), intravalley (blue), intervalley (green) and intervalley/($N_v$$-$1) (magenta) components versus energy (CBM is set to zero). The chemical potential corresponding to an electron density of 5$\times$10$^{19}$\,cm$^{-3}$ is indicated with a vertical line ($\mu$$-$$E_c$\,=\,117\,meV, 132\,meV, 122\,meV, $-$17\,meV, $-$14\,meV, 9\,meV for PbS, PbSe, PbTe, ScNiBi, ScPdSb and ZrNiSn). $T$\,=\,300\,K.} \label{fig:scattering_rate}
\end{figure*}

\subsection{Electron and phonon dispersions}
The electronic band structures for the PbX and HH materials are shown in Figs.~\ref{fig:bands_PbX}d)-f) and Figs.~\ref{fig:bands_HH}d)-f), respectively, along the high-symmetry points indicated in Fig.~\ref{fig:bands_PbX}b) and Fig.~\ref{fig:bands_HH}b). The energies are shifted such that the conduction band minimum (CBM) corresponds to zero. The PbX materials are direct gap semiconductors with the conduction and valence band edges located at the L point. This results in the conduction states having four equivalent valleys, $N_v$\,=\,4, as seen from the Fermi surface presented in Fig.~\ref{fig:bands_PbX}c). The calculated band gaps for PbS, PbSe, and PbTe are 0.51\,eV, 0.49\,eV, and 0.80\,eV, respectively. The corresponding experimental band gaps are 0.41\,eV, 0.28\,eV, and 0.31\,eV \cite{Skelton2014}.

The HH materials are indirect gap semiconductors with the conduction and valence band edges located at the X and $\Gamma$ points, respectively. As a result, the conduction band has three-fold degeneracy, $N_v$\,=\,3, as confirmed from the iso-energy contour plot shown in Fig.~\ref{fig:bands_HH}c). The band gaps obtained from DFT for ScNiBi, ScPdSb, and ZrNiSn are 0.24\,eV, 0.34\,eV, and 0.52\,eV, respectively. The experimental values for ScPdSb and ZrNiSn are 0.23\,eV and 0.13\,eV \cite{Oestreich2003,Schmitt2015}, with an estimated band gap of 0.084\,eV for ScNiBi from resistivity measurements \cite{Deng2017}. Comparing the PbX and HH materials, aside from the different number of equivalent conduction valleys mentioned above, another important distinction is that the more shallow conduction bands of the HH lead to significantly larger density of states. Important features shared by all the materials under study are that they possess multiple equivalent conduction valleys, and that their secondary bands have enough energy that they do not significantly participate in conduction ($\sim$\,10\,$k_B T$ above the CBM). This type of band structure naturally includes both intravalley and intervalley scattering, with the latter only arising from processes between one equivalent valley to another and no processes involving secondary bands that are out of range. This provides a convenient test case to characterize intravalley and intervalley el-ph scattering, which can easily be identified and separated depending on the phonon $|\mathbf{q}|$ involved. The general features of the electronic band structures of PbX and HH are consistent with the available literature \cite{Hummer2007,Guo2016,Winiarski2018,Zhou2018}.

The (unscreened) phonon dispersions for the PbX and HH materials are shown in Figs.~\ref{fig:bands_PbX}g)-i) and Figs.~\ref{fig:bands_HH}g)-i), respectively, along the same high-symmetry lines. With two atoms per primitive cell the PbX materials have six phonon branches, while there are nine phonon branches with the HH materials that have three atoms per primitive cell. In the PbX materials the maximum phonon energies vary from 27\,meV down to 14\,meV, which decreases with increasing atomic number of the chalcogen element. The zone-edge acoustic and optical phonons are as low as roughly 5\,meV; as discussed later this is relevant to intervalley scattering, particularly for the phonons near the X point. Another feature common to all PbX materials is a dip in the lowest optical phonon at the zone center, indicating that the rock-salt structure is energetically close to a phase transition to the rhombohedral structure \cite{Cao2021,Kilian2009}.

The HH materials display nearly the same maximum phonon energies around 30\,meV. Their zone-edge phonons, in particular those near the X point that result in intervalley scattering, have energies starting at 10\,meV, which is higher than those of the PbX materials. A feature of the HH phonon dispersions is the presence of energy gaps that originate from the large difference in the constituent masses \cite{Bano2018}. All the materials display a splitting of the optical phonons near the $\Gamma$ point due to the polar interaction, which are largely removed with the inclusion of carrier screening at an electron density of 5$\times$10$^{19}$\,cm$^{-3}$. (Our DFPT-calculated dielectric constants for PbS, PbSe, PbTe, ScNiBi, ScPdSb and ZrNiSn are 17.0, 20.8, 28.1, 26.3, 18.8 and 20.9, respectively, which are in agreement with published values \cite{Burstein1968,Roy2012}.) Overall, our phonon dispersions are consistent with that of others \cite{Skelton2014,Bano2018,Kocak2018,Ren2020}. Lastly, to test the accuracy of the electron and phonon dispersions calculated with the EPW code, Figs.~\ref{fig:bands_PbX}-\ref{fig:bands_HH} present excellent agreement between the dispersions obtained from QE (solid red lines) and EPW (dashed blue lines).

\begin{figure*}[ht]
\includegraphics[width=6.5in]{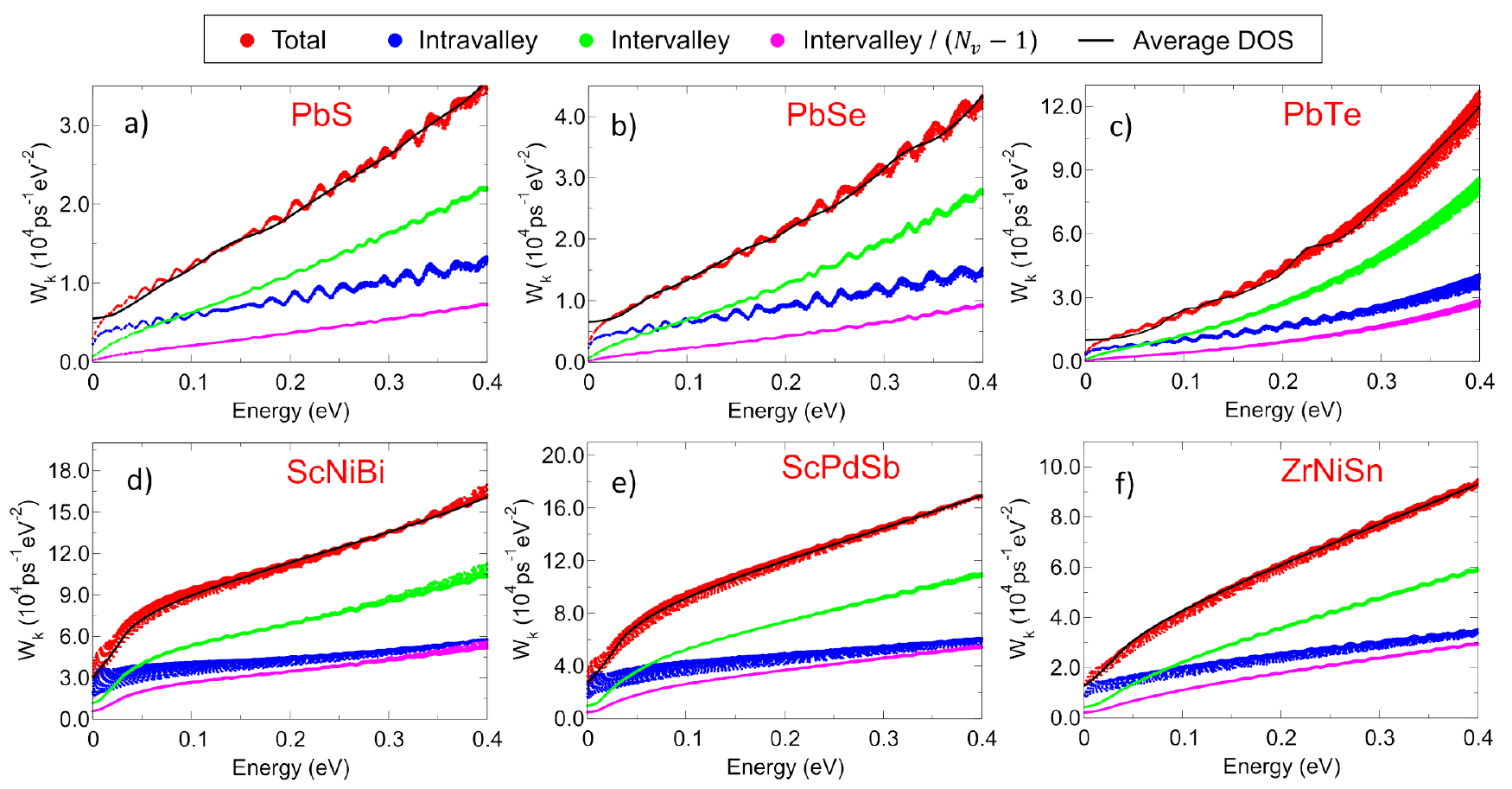}
\caption{Phase space, $W_{\bf k}$, separated into total (red), intravalley (blue), intervalley (green) and intervalley/($N_v$$-$1) (magenta) components versus energy (CBM is set to zero). Calculations are carried out for $T$\,=\,300\,K and an electron density of 5$\times$10$^{19}$\,cm$^{-3}$.} \label{fig:phase_space}
\end{figure*}

\subsection{Electron-phonon scattering rates}
Figure~\ref{fig:scattering_rate} presents the el-ph momentum scattering rates versus conduction band energy of all PbX (panels a)-c)) and HH materials (panels d)-f)), for an electron concentration of 5$\times$10$^{19}$\,cm$^{-3}$ and a temperature of 300\,K. The scattering contributions from intravalley and intervalley processes are shown separately, along with the intervalley rates divided by the number of final valleys ($N_v$\,$-$\,$1$). The latter allows for a comparison between intravalley and intervalley scattering on a per final valley basis, which is useful since the intervalley collisions would be expected to scale with the number of final valleys.

A general feature of the scattering rates is that they increase in energy away from the CBM. Qualitatively, this occurs since scattering scales with the number of final states, and thus generally follows the density of states which increases with energy. Focusing initially on the PbX materials, near the CBM the intravalley rates increase more rapidly than the intervalley rates, leading to overall higher contributions from intravalley scattering over the relevant energy range for carrier transport (roughly $\pm$10\,$k_B T$ around the chemical potential). This suggests that the PbX materials are good candidates for band convergence, from an intrinsic scattering perspective. The difference between intravalley and intervalley collisions grows further when comparing them on a per final valley basis (corresponding to dividing the intervalley rates by $N_v$\,$-$\,1\,=\,3). The PbX materials all display similar scattering rates below roughly 0.2\,eV, while at higher energies PbTe has the largest collision rates. With the HH materials, one can more clearly observe the onset of optical phonon emission scattering near 30\,meV resulting in a noticeable increase in the el-ph collision rates, particularly with ScNiBi. The intravalley and intervalley components are more similar compared to the PbX materials, and only when comparing both on a per valley basis (dividing the intervalley rates by $N_v$\,$-$\,1\,=\,2) does intravalley scattering appear slightly larger. Among the HH materials, ZrNiSn and ScPdSb display the lowest and highest collision rates, respectively.

Having highlighted the general features of the el-ph scattering rates, next we analyze their properties in further detail. In particular, we seek to understand what controls the degree of intravalley and intervalley processes, and how do these factors compare among the studied PbX and HH materials. To help address these questions and gain new insights, we rewrite the expression for the momentum scattering rate (Eq.~(\ref{eq:itau})) as the product of two terms:
\begin{align}
\tau_{\bf k}^{-1} =  W_{\bf k} \, \langle g^{2}_{\bf k} \rangle, \label{eq:itau2} 
\end{align}
where $W_{\bf k}$ is the el-ph scattering phase space and $\langle g^{2}_{\bf k} \rangle$ is the average el-ph coupling squared. These two quantities are defined as:
\begin{align}
W_{\bf k}  =& \sum_{\bf k'}  W_{{\bf k},{\bf k'}}, \label{eq:wk} \\
 \langle g^{2}_{\bf k} \rangle  =& \frac{ \sum_{\bf k'}  |g({\bf k},{\bf k'})|^2 W_{{\bf k},{\bf k'}} }{ \sum_{\bf k'}  W_{{\bf k},{\bf k'}} } , \label{eq:avg} 
\end{align}
in terms of $W_{{\bf k},{\bf k'}}$ given by:
\begin{align}
 W_{{\bf k},{\bf k'}} =& \frac{2\pi}{\hbar} \big[ n_{\bf q} + 1/2 \pm (f_{\bf k'} - 1/2) \big] \delta_{{\bf k'},{\bf k}\pm{\bf q}} \nonumber \\
& \delta(E_{\bf k}-E_{\bf k'} \pm \hbar \omega_{\bf q})    \Big( 1-\frac{{\bf v}_{\bf k'}\cdot{\bf v}_{\bf k}} {|{\bf v}_{\bf k'}|{|\bf v}_{\bf k}|}  \Big). \label{eq:wkk}
 \end{align}
The explicit conservation of crystal momentum is introduced into Eq.~(\ref{eq:wkk}), which is normally enforced via the el-ph coupling factor $g({\bf k},{\bf k'})$. The phase space $W_{\bf k}$ scales with the number of possible el-ph transitions given the constraints of crystal momentum and energy conservation, and only depends on electron and phonon dispersions. This quantity is closely related to its counterpart describing three-phonon scattering \cite{Lindsay2008,Li2014,Strongman2021}. The el-ph interaction strength is captured by the average el-ph coupling squared $\langle g^{2}_{\bf k} \rangle$, which reduces the dimensionality of the full el-ph coupling matrix $g({\bf k},{\bf k'})$ in order to faciliate its analysis and interpretion. Loosely speaking, for a given electron state $\bf k$, the phase space $W_{\bf k}$ indicates how many scattering transitions are possible and the average coupling $\langle g^{2}_{\bf k} \rangle$ reflects how likely those collision processes are to occur. Next, we will compare both these quantities for all the PbX and HH materials.

\begin{figure*}[ht]
\includegraphics[width=6.5in]{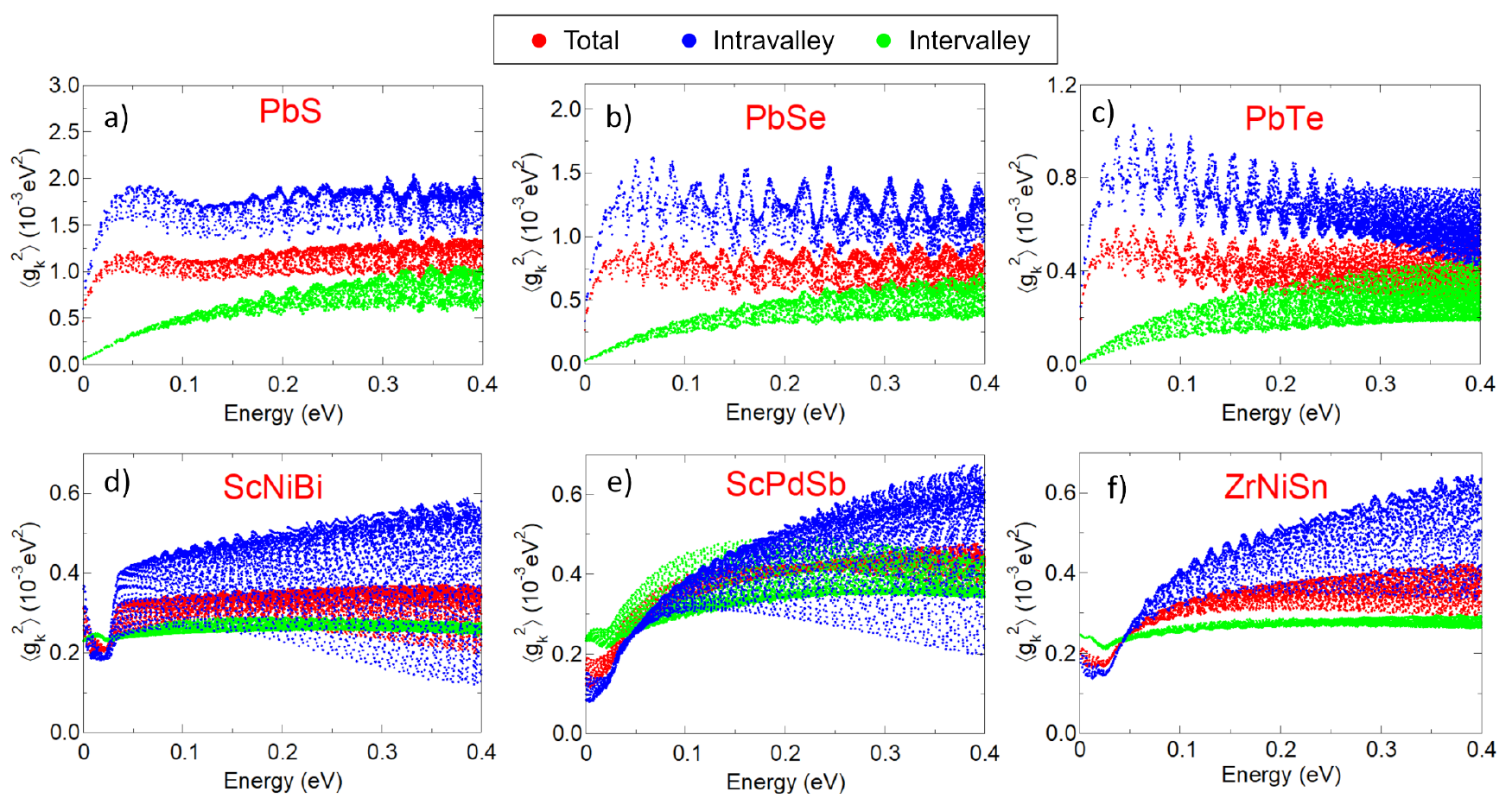}
\caption{Average electron-phonon coupling squared, $\langle g^2_{\bf k} \rangle$, separated into total (red), intravalley (blue) and intervalley (green) components versus energy (CBM is set to zero). Calculations are carried out for $T$\,=\,300\,K and an electron density of 5$\times$10$^{19}$\,cm$^{-3}$.} \label{fig:g2}
\end{figure*}

\subsection{Phase space}
Figure~\ref{fig:phase_space} presents the scattering phase space versus energy for all the PbX and HH materials, with individual components due to intravalley and intervalley processes. The energy dependence of the phase space is closely related to that of the final average electron DOS. This is illustrated by comparing the total $W_{\bf k}$ (red markers) to the energy-averaged DOS (black lines), $\bar{D}(E)$\,=\,$\int_{E-\hbar\omega_{\rm max}}^{E+\hbar\omega_{\rm max}} D(E')\,dE'/(2\hbar\omega_{\rm max})$, where $\hbar\omega_{\rm max}$ is the maximum phonon energy and the DOS averaging occurs over the possible energy range of final states. The excellent agreement between $W_{\bf k}$ and $\bar{D}(E)$ indicates that the phase space is largely controlled by the DOS, which is easy to interpret and compute. This explains why the PbX materials display much lower phase space compared to the HH materials -- the lower DOS of the PbX materials results in fewer possible el-ph transitions -- and describes the trend within each material class.

It is interesting to compare $W_{\bf k}^{\rm intra}$ and $W_{\bf k}^{\rm inter}$/($N_v$\,$-$\,1), because both are expected to scale with the DOS of an individual valley. And while both these quantities run parallel to each other, the intravalley component is consistently larger. The reason for this can be understood from the definition of $W_{\bf k}$ given by Eqns.~(\ref{eq:wk})-(\ref{eq:wkk}), which shows that the phase space is proportional to the equilibrium occupation of the phonons participating in the scattering, $n_{\bf q}$. Phonons with larger energy have lower occupation, and vice versa. Intravalley processes involve small ${\bf q}$ wavevector phonons, such as acoustic phonons, that tend to have small energies and large $n_{\bf q}$ when compared to the large ${\bf q}$ phonons required for intervalley transitions (in the case of PbX and HH materials, intervalley processes involve phonons near the X point at the zone edge). Because of this difference in phonon energies, $W_{\bf k}^{\rm intra}$ is larger than $W_{\bf k}^{\rm inter}$/($N_v$\,$-$\,1) for all the studied materials.

\begin{figure*}[ht]
\includegraphics[width=5.5in]{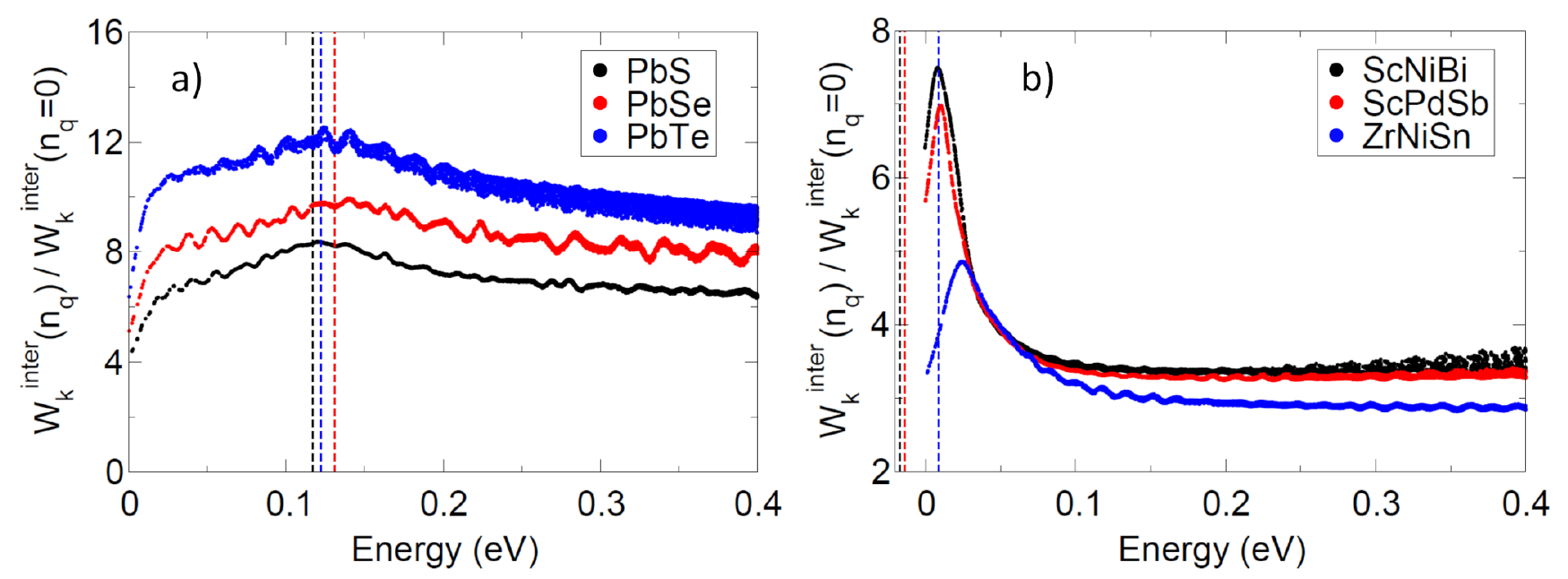}
\caption{Ratio of intervalley phase space with and without setting the phonon occupation to zero, $W_{\bf k}^{\rm inter}(n_{\bf q})/W_{\bf k}^{\rm inter}(n_{\bf q}$\,=\,0), versus energy (CBM is set to zero). Calculations are carried out for $T$\,=\,300\,K and an electron density of 5$\times$10$^{19}$\,cm$^{-3}$. The chemical potential is indicated with a vertical dashed line.} \label{fig:W_ratio}
\end{figure*}

\subsection{Average el-ph coupling squared}
Next, we focus on the contribution of the average el-ph coupling squared ${\langle}g^{2}_{\bf k}{\rangle}$ to the scattering rates, which are presented in Fig.~\ref{fig:g2}. We begin by noting how the total ${\langle}g^{2}_{\bf k}{\rangle}$ (red markers) are relatively independent of energy for all six materials, but particularly with PbX -- this type of observation can be useful when seeking to model the scattering properties in a simple approximate way. The breakdown of ${\langle}g^{2}_{\bf k}{\rangle}$ into intravalley and intervalley components shows that the former is significantly larger with the PbX materials. One reason for this is that the small ${\bf q}$ phonons involved with intravalley transitions tend to have stronger long-range polar interaction compared to the large ${\bf q}$ phonons responsible for intervalley scattering. Another reason for the disparity between ${\langle}g^{2}_{\bf k}{\rangle}^{\rm intra}$ and ${\langle}g^{2}_{\bf k}{\rangle}^{\rm inter}$ is that with the PbX materials scattering between the CBM of different valleys is prohibited due to symmetry \cite{Cao2018}, which explains why the intervalley component vanishes as the energy approaches the CBM. The el-ph coupling strength decreases in PbX with a larger chalcogen atom, since those elements have reduced electronegativity, smaller Born effective charges and weaker polar interaction. With the HH materials ${\langle}g^{2}_{\bf k}{\rangle}^{\rm intra}$ is only slightly larger than ${\langle}g^{2}_{\bf k}{\rangle}^{\rm inter}$, with ScPdSb showing nearly equal contributions. One interesting feature of the HH materials is their overall low ${\langle}g^{2}_{\bf k}{\rangle}$ compared to the PbX materials, which is attributed to the non-bonding character of the conduction states \cite{Zhou2018}.

Note that the total ${\langle}g^{2}_{\bf k}{\rangle}$ is not equal to the sum of the intravalley and intervalley contributions. Using the definitions $1/\tau_{\bf k}^{\rm intra}$\,=\,$W_{\bf k}^{\rm intra} \langle g^{2}_{\bf k} \rangle^{\rm intra}$ and $1/\tau_{\bf k}^{\rm inter}$\,=\,$W_{\bf k}^{\rm inter} \langle g^{2}_{\bf k} \rangle^{\rm inter}$, along with $1/\tau_{\bf k}$ = $1/\tau_{\bf k}^{\rm intra}$ + $1/\tau_{\bf k}^{\rm inter}$, we arrive at the relation:
\begin{align}
 \langle g^{2}_{\bf k} \rangle  &= \frac{ W_{\bf k}^{\rm intra} \langle g^{2}_{\bf k} \rangle^{\rm intra} + W_{\bf k}^{\rm inter} \langle g^{2}_{\bf k} \rangle^{\rm inter} }{ W_{\bf k}^{\rm intra} + W_{\bf k}^{\rm inter} }, \label{eq:avg2}
\end{align}
where $W_{\bf k}$\,=\,$W_{\bf k}^{\rm intra}$\,+\,$W_{\bf k}^{\rm inter}$. Equation~(\ref{eq:avg2}) shows that the average el-ph coupling depends on the intravalley and intervalleys components weighted by their respective phase space component.

\section{Discussion} 
\label{sec:discussion}
Decomposing the scattering rates into the phase space and average el-ph coupling helps understand the origin of the scattering characteristics. As shown above, the phase space -- a measure of how many el-ph transitions are possible -- is largely determined by the density of states (or equivalently in terms of effective mass), which can be easily calculated for different materials. When the average el-ph coupling is energy independent, which appears to be a reasonable starting assumption based on the calculated $\langle g^2_{\bf k} \rangle$ in Fig.~\ref{fig:g2}, the scattering rates are simply related to the density of states. This helps explain why DFT-calculated el-ph scattering rates have been shown to agree with the DOS \cite{Witkoske2017,Askarpour2019}.

It is also interesting to note that while the scattering rates are fairly similar among both classes of PbX and HH materials (in the range 15-25\,ps$^{-1}$ for 0.1\,eV above the CBM), our analysis of the phase space and average el-ph coupling indicates that this arises for different reasons; the HH materials display lower $\langle g^2_{\bf k} \rangle$ and higher $W_{\bf k}$ compared to the PbX materials in which this trend is reversed.

Next, we explore how one could search for or design materials with limited intervalley collisions, leading to improved thermoelectric power factor and $ZT$. Equation~(\ref{eq:itau2}) informs us that a reduction in intervalley scattering can originate from the phase space or the average el-ph coupling. Focusing first on the phase space, Eqns.~(\ref{eq:wk}),(\ref{eq:wkk}) show that this quantity is proportional to the equilibrium occupation of the phonons participating in the scattering, $n_{\bf q}$, which for the PbX and HH materials correspond to zone-boundary phonons near $\bf q$\,$\approx$\,X. This suggests that materials with high-energy phonons, specifically the phonons responsible for the electronic transitions from one valley to another, can lead to reduced $W_{\bf k}^{\rm inter}$ and $1/\tau_{\bf k}^{\rm inter}$. The Bose-Einstein distribution $n_{\bf q}$ monotonically decreases with increasing energy, and for energies beyond a few $k_B T$ decays exponentially -- thus, this approach would be expected to be most effective with relatively large phonon energies or at lower temperatures. Such a strategy has been reported in cubic boron-V compounds such as BP, BAs and BSb \cite{Liu2018}.

To explore the potential reduction in intervalley scattering, we calculate the phase space for intervalley transitions assuming that $n_{\bf q}$\,=\,0, which represents a hypothetical limit where the energy of the zone-edge phonons is very large, $\hbar \omega_{\bf q}$\,$\gg$\,$k_B T$. Figure~\ref{fig:W_ratio} presents the ratio of the intervalley phase space with and without setting $n_{\bf q}$\,=\,0. The PbX and HH materials show a possible reduction in $W_{\bf k}^{\rm inter}$ reaching between 8-12 and 5-7.5, respectively, if the scattering-mediated phonons had negligible occupation. With the PbX materials, the effect increases going from PbS to PbSe to PbTe, since the lowest-energy phonon at $\bf q$\,=\,X decreases in energy (increasing $n_{\bf q}$) among these materials. The same argument explains the increasing phase space ratio of the HH materials going from ZrNiSn to ScPdSb to ScNiBi. The phase space ratio has a different shape for both material classes, which is due to the location of the chemical potential; a $\mu$ located deeper in the band (PbX) allows for both absorption and emission processes around $\mu$ giving rise to a peak, while a $\mu$ below the band edge (HH) predominantly has absorption-only processes near $\mu$ giving rise to a sharp peak and rapid drop.

\begin{figure*}[ht]
\includegraphics[width=6.5in]{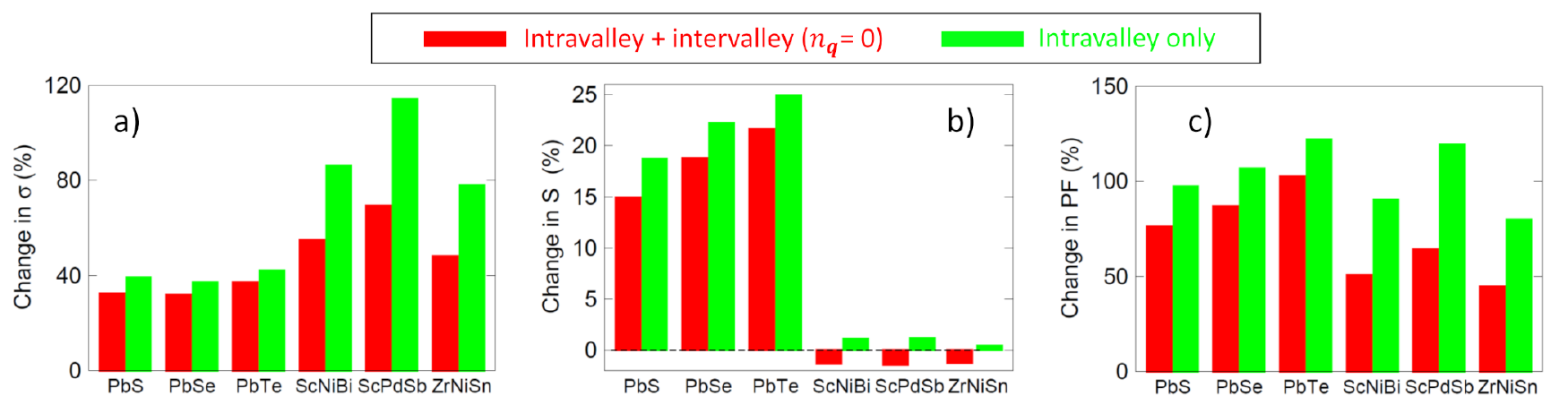}
\caption{Change in a) conductivity, b) Seebeck coefficient, and c) power factor when setting $n_{\bf q}$\,=\,0 for intervalley processes (red) and only including intravalley transitions (green). Calculations are carried out for $T$\,=\,300\,K and an electron density of 5$\times$10$^{19}$\,cm$^{-3}$.} \label{fig:TE}
\end{figure*}

Figure~\ref{fig:TE} shows how the thermoelectric parameters change when $n_{\bf q}$ is set to zero (for intervalley processes only), in order to explore how this could improve the TE properties. For comparison, we also present the change in TE coefficients when eliminating intervalley transitions entirely, leaving only intravalley scattering. (The original TE properties, including all intravalley and intervalley processes, are presented in Appendix~\ref{app:thermoelectric}.) The power factor increase is in the range of 45-100\% among the different semiconductors. With the PbX materials this comes from a combination of enhanced electrical conductivity and Seebeck coefficient, while with the HH materials this is attributed to only the electrical conductivity. The increase in electrical conductivity, corresponding to $\sim$35\% with the PbX and 50-70\% with the HH, simply results from reduced el-ph scattering. This effect is most pronounced in the HH materials since intervalley collisions represent a larger fraction of the overall scattering compared to the PbX materials. By only considering intravalley scattering, the conductivity of the PbX and HH semiconductors increases further by $\sim$5\% and 30-45\%, respectively. The Seebeck coefficient is related to the average energy of electron flow relative to the chemical potential, $\langle E  - \mu \rangle$. Suppressing intervalley scattering can alter the energy dependence of el-ph scattering profile, and in the case of the PbX materials can increase the lifetime of higher-energy electrons thereby increasing $\langle E  - \mu \rangle$ and $|S|$. These results highlight the potential benefits of materials with larger-energy zone-edge phonons to reduce intervalley scattering. Moreover, Fig.~\ref{fig:TE} demonstrates that not considering intervalley scattering in theoretical calculations can significantly overestimate the conductivity and power factor.

\begin{figure*}[ht]
\includegraphics[width=5.0in]{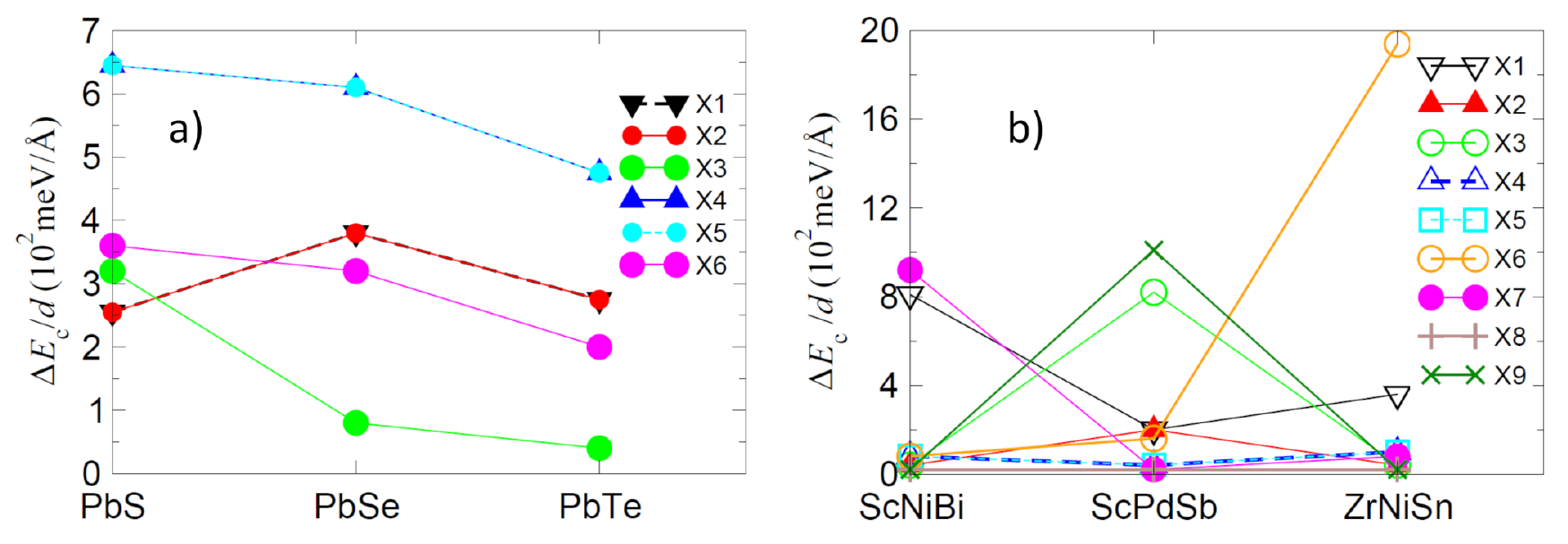}
\caption{Average magnitude shift in CBM energy divided by maximum atomic displacement, $\Delta E_c / d$, in a 1$\times$2$\times$2 supercell simulating the different vibrational modes of the ${\bf q}$\,=\,X zone-edge phonons. The CBM variations are calculated relative to the unperturbed supercell. The phonon modes are labelled from lowest to highest energy.}\label{fig:cbm_shift}
\end{figure*}

\begin{figure*}[ht]
\includegraphics[width=6.5in]{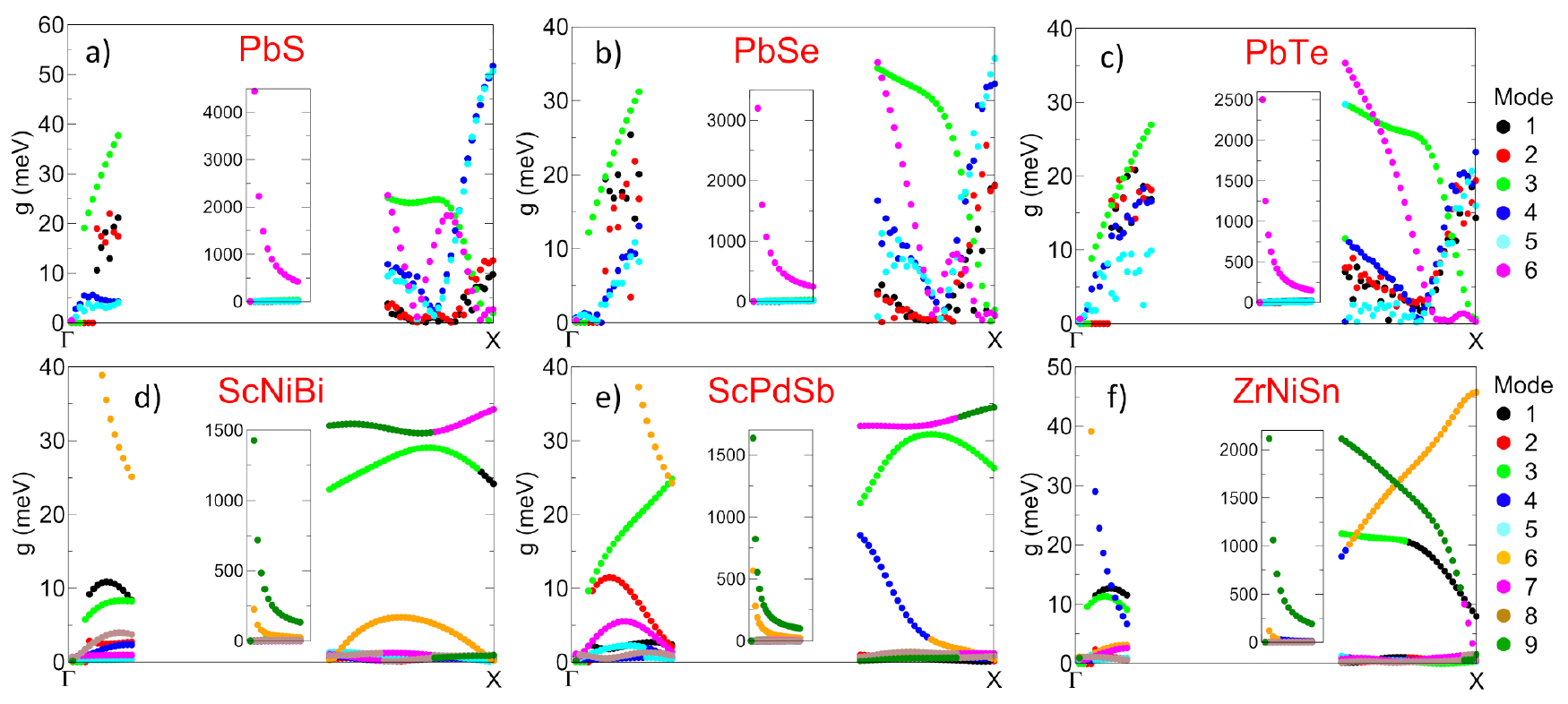}
\caption{Mode-resolved electron-phonon coupling (unscreened), $g({\bf k},{\bf q})$, versus phonon wavevector $\bf q$ along the $\Gamma$-X direction for an initial electron state $\bf k$ located 0.1\,eV above the CBM along the $\Gamma$-X line. The insets show the large contributions arising near the zone center due to the polar interaction. The phonon modes are labelled from lowest to highest energy.} \label{fig:gkk_polar}
\end{figure*}

So far we have focused on how the intervalley phase space $W_{\bf k}^{\rm inter}$ could be suppressed, however the average el-ph coupling $\langle g^2_{\bf k} \rangle^{\rm inter}$ is equally important. Unlike the phase space that can be reasonably approximated by the DOS, an easily computed quantity, obtaining the average el-ph coupling for intervalley processes requires a rigorous calculation that is computationally demanding. Here, we propose a simple and efficient method for estimating $\langle g^2_{\bf k} \rangle^{\rm inter}$ that can be used, for example, to screen through material candidates.

Inspired by the deformation potential approximation \cite{Bardeen1950,Herring1956}, this approach relies on constructing a supercell in which the atoms are displaced according to the eigenvectors of the zone-edge phonons participating in the intervalley processes and calculating the change in conduction band edge energy $\Delta E_c$. Since intervalley transitions typically involve zone-boundary phonons with short wavelengths, the supercells can be relatively compact -- for the PbX and HH materials studied in this work, the atomic displacements of the $\bf q$\,=\,X phonons responsible for intervalley processes can be directly simulated with a 1$\times$2$\times$2 supercell (see Appendix~\ref{app:supercell} for details on how to construct the supercell). The normalized eigendisplacements are scaled such that the maximum atomic displacement is $d$\,=\,0.05\,\AA. Since $\Delta E_c$ is expected to scale with $d$, we quote $\Delta E_c/d$ as a measure of the intervalley el-ph coupling strength. One supercell calculation is carried out for each of the zone-edge phonon modes (six and nine in the case of PbX and HH, respectively). The atomic perturbation in the supercell alters the energy and degeneracy of the band structure, and $\Delta E_c$ is calculated as the average magnitude in CBM shift relative to the unperturbed supercell. Note that this supercell approach is only meant to provide a qualitative measure of the intervalley el-ph coupling, which can potentially serve to compare and assess different materials.

Figure~\ref{fig:cbm_shift} presents $\Delta E_c/d$ for the PbX and HH semiconductors. With the PbX materials, the largest values correspond to the phonon modes 4-5 and increase going from PbTe to PbSe to PbS. This trend is consistent with the calculated $\langle g^2_{\bf k} \rangle^{\rm inter}$ shown in Fig.~\ref{fig:g2}. In the case of the HH materials, the largest $\Delta E_c/d$ values among the different phonon modes increases going from ScNiBi to ScPdSb to ZrNiSn. This is somewhat different from the $\langle g^2_{\bf k} \rangle^{\rm inter}$ in Fig.~\ref{fig:g2} showing that the average intervalley el-ph coupling is highest in ScPdSb and nearly identical in ScNiBi and ZrNiSn. Comparing the maximum $\Delta E_c/d$ among both material classes, the HH semiconductors have larger values than the PbX semiconductors. While the HH materials displayed lower overall $\langle g^2_{\bf k} \rangle$ in Fig.~\ref{fig:g2}, which was explained in terms of non-bonding orbital hybridization of the conduction states \cite{Zhou2018}, the intervalley component of $\langle g^2_{\bf k} \rangle$ is lower in the PbX materials for energies near the band edge.

This supercell approach can potentially identify which phonon branches control the intervalley el-ph coupling. To verify that the results of this simple method are qualitatively consistent with those from DFT, we calculate the mode-resolved el-ph coupling $g({\bf k},{\bf q})$ versus phonon wavevector $\bf q$ along the line $\Gamma$-X, as shown in Fig.~\ref{fig:gkk_polar}. The initial electron state $\bf k$ is located 0.1\,eV above the CBM along the $\Gamma$-X line. Note that there is a gap in $g({\bf k},{\bf q})$ for intermediate $q$ values, which separates the intravalley (small $q$) and intervalley (large $q$) processes. Since the supercell approach captures the effect of phonons at the $\bf q$\,=\,X point, we focus on the results of $g({\bf k},{\bf q}$\,=\,X) for comparison.

With the PbX semiconductors, Fig.~\ref{fig:gkk_polar} shows that modes 4-5 are dominant and that they increase when going from PbTe to PbSe to PbS, which is consistent with findings of the supercell approach (Fig.~\ref{fig:cbm_shift}). With the HH semiconductors, Fig.~\ref{fig:gkk_polar} indicates that the two largest contributions come from modes 7 and 1 in ScNiBi, 9 and 3 in ScPdSb, and 6 and 1 in ZrNiSn -- the supercell approach identified the same two dominant phonon modes for each HH (Fig.~\ref{fig:cbm_shift}). The maximum $g({\bf k},{\bf q}$\,=\,X) is found in ZrNiSn, with ScNiBi and ScPdSb having lower values that are similar, which is in qualitative agreement with the results of Fig.~\ref{fig:cbm_shift}. These findings suggest that this simple and efficient supercell approach may be useful to identify materials with low intervalley el-ph coupling when searching for new TE materials.

\section{Conclusions}
\label{sec:conclusion}
Intervalley collisions in semiconductors comprised of multiple valleys/bands can significantly deteriorate thermoelectric performance, through enhanced scattering that lowers the electrical conductivity and power factor. In this work, first-principles electron-phonon scattering and electron transport calculations were carried out in three lead chalcogenides (PbS, PbSe, PbTe) and three half-Heuslers (ScNiBi, ScPdSb, ZrNiSn), which all possess multiple equivalent conduction valleys giving rise to intervalley scattering. The individual intravalley and intervalley el-ph scattering components were isolated and compared. The results show that intervalley collisions represent a smaller fraction of the total el-ph scattering in the lead chalcogenides (PbX) compared to the half-Heuslers (HH), and that the intervalley scattering rates scale with the valley degeneracy leading to stronger energy dependence versus intravalley processes.

To better understand what controls the scattering characteristics, the collision rates are expressed as the product of the phase space $W_{\bf k}$, a measure of how much scattering is possible given the constraints of conservation of energy and crystal momentum, and the average el-ph coupling squared $\langle g^2_{\bf k} \rangle$, a measure of the probability of transitions occurring. The phase space is found to follow the electronic density of states, which can easily be computed, and explains why the HH materials show larger $W_{\bf k}$ over the PbX materials. The average el-ph coupling for intravelley transitions is significantly larger than intervalley transitions with the PbX materials, due to a stronger polar interaction for the small $\bf q$ phonons participating in the intravalley processes and prohibited intervalley processes at the conduction band minimum. The HH materials displayed more similar $\langle g^2_{\bf k} \rangle$ for both intravalley and intervalley, however the overall el-ph coupling is weaker compared to the PbX materials as a result of their particular bonding properties. While the el-ph scattering rates are somewhat similar for both the PbX and HH semiconductors, an analysis of their phase space and average el-ph coupling reveal different underlying characteristics.

Lastly, we explored two approaches that could help guide the search for new and improved thermoelectric materials with reduced intervalley scattering. First, increasing the energy of the zone-edge phonons responsible for the intervalley processes lowers the transition rate. In the limit when $\hbar\omega$\,$\gg$\,$k_BT$, we find that the intervalley phase space can drop by an order of magnitude and the power factor can double. Second, we introduce a simple and efficient supercell approach to estimate the intervalley el-ph coupling strength arising from zone-boundary phonons. When compared to rigorous DFT-computed el-ph coupling, this supercell method is found to be in good qualitative agreement, and correctly identified trends among the different materials as well as the two dominant phonon modes in each case. The findings of this study help advance our understanding of what controls intravalley and intervalley el-ph scattering, which, coupled with the proposed strategies to identify and design materials with suppressed intervalley scattering, may lead to the discovery of new high-performance thermoelectrics and high-mobility semiconductors.

\section*{Acknowledgement}
This work was supported by NSERC (Discovery Grant RGPIN-2016-04881) with computing resources provided by the Digital Research Alliance of Canada.

\appendix
\section{Thermoelectric transport parameters}
\label{app:thermoelectric}
Table~\ref{tab:te} presents the calculated electrical conductivity $\sigma$, Seebeck coefficient $S$ and power factor $PF$\,=\,$S^2 \sigma$ for the three lead chalcogenides and the three half-Heuslers. Calculations are carried out at 300\,K with an electron density of 5$\times$10$^{19}$\,cm$^{-3}$, and include all intravalley and intervalley processes.

\begin{table}
	\begin{tabular}{|c || c | c | c |} 
		\hline
		 & $\sigma$ (S/m) & $S$ ($\mu$V/K) & $PF$ ($\mu$W/cm-K$^2$)\\ [0.5ex] 
		\hline\hline
		PbS  & 4.95$\times$10$^5$ & -55.6 & 15.3 \\ 
		\hline
		PbSe & 5.67$\times$10$^5$ & -49.5 & 13.9 \\ 
		\hline
		PbTe & 5.38$\times$10$^5$ & -51.8 & 14.4 \\ 
		\hline
		ScNiBi & 1.07$\times$10$^5$ & -228 & 55.6 \\ 
		\hline
		ScPdSb & 1.10$\times$10$^5$ & -207 & 47.1 \\ 
		\hline
		ZrNiSn & 3.15$\times$10$^5$ & -165 & 85.8 \\ 
		\hline
	\end{tabular}
	\caption{Thermoelectric transport coefficients calculated at 300\,K and an electron density of 5$\times$10$^{19}$\,cm$^{-3}$, including all intravalley and intervalley processes.}
	\label{tab:te}
\end{table}

\section{Supercell for estimating intervalley electron-phonon coupling}
\label{app:supercell}
What is the smallest supercell needed to capture the atomic displacement of a zone-boundary phonon? The atomic displacement associated with a particular phonon mode is given as the product of the cell-periodic normalized phonon eigenvector and a phase factor of the form $\exp(i{\bf q}\cdot{\bf R})$, where 
\begin{align}
{\bf R} &= \sum_{\alpha} n_{\alpha} {\bf a}_{\alpha} = n_1 {\bf a}_1 +  n_2 {\bf a}_2 + n_3 {\bf a}_3, \label{eq:lattvec}
\end{align}
${\bf a}$ are the primitive lattice vectors, $n_{\alpha}$ are integers, and $\bf q$ is the phonon wavevector. Since the eigenvectors share the same periodicity as the primitive cell, the supercell size will be determined by the phase factor. Next, we express the phonon wavevector as 
\begin{align}
{\bf q} &= \sum_{\beta} \nu_{\beta} {\bf b}_{\beta} = \nu_1 {\bf b}_1 +  \nu_2 {\bf b}_2 + \nu_3 {\bf b}_3, \label{eq:recivec}
\end{align}
where ${\bf b}$ are the reciproval lattice vectors, and $\nu_{\beta}$ are real numbers (here assumed to restrict ${\bf q}$ to within the Brillouin zone). The phase factor can then be written as
\begin{align}
\exp(i{\bf q}\cdot{\bf R}) &= \exp\Big( i \sum_{\alpha,\beta} n_{\alpha} \nu_{\beta} \, {\bf a}_{\alpha}\cdot{\bf b}_{\beta} \Big) \nonumber \\
&= \exp\big[ i 2\pi (n_1 \nu_1 + n_2 \nu_2 + n_3 \nu_3) \big], \label{eq:phasefactor1}
\end{align}
where we used the property ${\bf a}_{\alpha}\cdot{\bf b}_{\beta}$\,=\,$2\pi\delta_{\alpha,\beta}$ that comes from the definition of the reciprocal lattice vectors. For a face-centered cubic crystal, the X point corresponds to ${\bf q}_{\rm X}$\,=\,$({\bf b}_2 + {\bf b}_3)/2$. Inserting this into Eq.~(\ref{eq:phasefactor1}), we find
\begin{align}
\exp(i{\bf q}_{\rm X}\cdot{\bf R}) &= \exp\big[ i \pi (n_2 + n_3) \big]. \label{eq:phasefactor2}
\end{align}
This indicates that the phase factor for a ${\bf q}_{\rm X}$ phonon is periodic with all even or odd integers of $n_2$ and $n_3$, such that the wavelength of the atomic displacements matches a 1$\times$2$\times$2 supercell. Thus, for a particular phonon mode at the X point, the atomic displacements in the 1$\times$2$\times$2 supercell are obtained by the product of the cell-periodic eigenvectors times the phase factor that loops over $n_{2,3}$\,=\,1, 2 (each of the four sub-cells that make up the supercell are multiplied by a different phase factor as determined by the value of $n_2$ and $n_3$).


\end{document}